\documentstyle[12pt]{article}

\newcommand{\be}{\begin{equation}}
\newcommand{\ee}{\end{equation}}

\def\psnormal{\textwidth=16cm\textheight=21.5cm
          \oddsidemargin=0.5cm\evensidemargin=0cm
          \topmargin=0cm\parindent=1cm}
\psnormal
\def\a{\alpha}
\def\P{\Phi}
\def\d{\delta}
\def\O{\Omega}
\def\G{\Gamma}

\begin{document}
\pagestyle{empty}

\hspace{3cm}

\vspace{-3.4cm}
{\vbox{\baselineskip=12pt{\rightline{\small{ CERN--TH.6958/93}}
\rightline{\small{NEIP--93--006}}}
\rightline{{\small{IEM--FT--75/93}}}}

\vspace{1cm}
\begin{center}
%
{\bf {\large M}ODEL-{\large I}NDEPENDENT {\large P}ROPERTIES AND\\
{\large
 C}OSMOLOGICAL
{\large I}MPLICATIONS
OF THE  {\large D}ILATON AND
 {\large M}ODULI\\
{\large S}ECTORS  OF {\large 4-D} {\large S}TRINGS}
%
\vspace{1cm}

B. de CARLOS${}^*$, \ J.A. CASAS${}^{**,*}\;$, \ F. QUEVEDO${}^{***}
\footnote{Supported by the Swiss National Science Foundation}\;$,
\  E. ROULET${}^{**}$
\vspace{0.7cm}

${}^{*}$ {\it Instituto de Estructura de la Materia (CSIC),\\
Serrano 123, 28006--Madrid, Spain}
\vspace{0.4cm}

${}^{**}$ {\it CERN, CH--1211 Geneva 23, Switzerland }
\vspace{0.3cm}

${}^{***}$ {\it Institut de Physique, Universit\'e de Neuch\^atel,\\
CH--2000 Neuch\^atel, Switzerland }
\vspace{0.3cm}

\end{center}

\vspace{0.7cm}

{\vbox{\baselineskip=15pt
\noindent
We show that if there is a realistic  4-d string, the dilaton and
moduli supermultiplets will generically acquire a small mass $\sim
{O}(m_{3/2})$,  providing the only vacuum-independent evidence of
low-energy physics  in string theory
beyond the supersymmetric standard model. The only assumptions behind
this result are
(i) softly broken supersymmetry at low energies with
zero cosmological constant,
(ii) these particles interact with gravitational strength and the
scalar  components have a  flat potential in perturbation theory,
which are well-known properties of string theories.
 (iii) They acquire a $vev$ of the order of the
Planck scale (as required
for the correct value of the gauge coupling constants and the expected
compactification scale) after supersymmetry gets broken.  We explore
the cosmological  implications of these  particles. Similar to the
gravitino, the fermionic states may overclose the Universe if they are
stable or destroy nucleosynthesis if they decay unless their  masses
belong to a certain range  or inflation dilutes them. For the scalar
states it is known that the problem cannot be entirely solved by
inflation, since
oscillations around the minimum of the potential, rather than thermal
production, are the main source for their energy and can lead to a huge
entropy generation at late times.
We discuss some possible ways to alleviate this entropy problem, that
favour low-temperature baryogenesis, and also
comment on the possible role of these  particles as dark matter
candidates or as sources of the baryon asymmetry through their decay.}}

\vspace{0.5cm}
\begin{flushleft}
{\vbox{\baselineskip=12pt {\small{CERN--TH.6958/93}} \\
{\small{July 1993}}}}
\end{flushleft}
\psnormal
%
%

\newpage
\baselineskip=16pt
\pagestyle{plain}
\pagenumbering{arabic}
\section{Introduction}

One of the major problems facing string theories is the lack
of predictive power for low-energy physics. This problem is mostly
due
to the immense number of consistent supersymmetric vacua that
can be constructed. 
Concentrating on  model-independent properties of these vacua, it is
well
known that all of them include in the spectrum, besides the gravity
sector, a dilaton ($S$) multiplet. 
The tree-level couplings of this field
are well understood and are independent of the vacuum. In particular
it couples to the gauge kinetic terms,  thus its $vev$ gives the gauge
 coupling constant at the string scale.
It is also known to have vanishing potential to all orders in
perturbation
theory and to couple to every other light field only by non-renormalizable
interactions.

Another generic property of 4-d string vacua is that each  of them
belongs to
 classes of models labelled by continuous parameters called moduli
($T_i$).
These parameters are fields in the effective field theory whose
potential is flat. The standard moduli fields 
 characterize
a particular compactification, e.g. the size and shape of a
Calabi--Yau
manifold or toroidal orbifold. But they are also known to exist even
in models without a clear geometric interpretation like asymmetric
orbifolds
\footnote{Examples of models without moduli have been constructed, but
none of them corresponds to a 4-d string vacuum \cite{HMV}.
}.
Even though their existence is generic, their couplings are not as
model-independent as those of the dilaton, except for having vanishing
potential
and non-renormalizable couplings to the other light fields.

Independent of the particular mechanism of  supersymmetry breaking,
any realistic string model is expected to lead to a low-energy theory
with softly
broken supersymmetry at low scale in order to solve the hierarchy
problem. Experimental constraints also impose  that the cosmological
 constant
is essentially zero and that the gauge coupling constant
at the unification scale is of
order one.
Also the size and shape (moduli) of the compact dimensions are expected
to be  of  order the Planck scale.

 These phenomenological requirements, together with the assumption that
the supersymmetry breaking mechanism is responsable for fixing those
$vev$'s, are
enough to prove a very simple but strong result in string theory:
The scalar and fermionic components of the dilaton and moduli
superfields can take masses at most of the order of the
 gravitino mass, which is expected to be $\leq 1$ TeV
in order to solve
the hierarchy problem. This means that any realistic string theory
(as defined
 above) should have at low-energies not only the supersymmetric standard
model spectrum, but also the dilaton and moduli superfields
\footnote{Strictly speaking, not all of the moduli fields have to
take
$vev$'s of order 1. Our result applies only to those that have a
$vev$ of that order, having geometrical interpretation or not.
The logical alternatives are that non-perturbative effects do not lift
 the degeneracy
completely, thus leaving some flat directions and then massless fields
or that the $vev$'s are smaller (larger) than the Planck scale for which
 the
corresponding
field is heavier (lighter). For an example in this class see \cite{AMQ}.}.
This fact is actually not surprising since before supersymmetry breaking
these fields have flat potentials (and are therefore massless).
The interesting observation is that this can be  claimed in a
completely vacuum-independent
way and is therefore a `prediction' of  any model satisfying the
above hypothesis, which  are almost imperative
in any possible realistic
string model (note, however, that in ordinary field theories the $S$
and $T_i$ fields are not necessarily present).

It is then natural to ask what  the physical implications of these
light particles are. Since all of their couplings to the observable sector
are suppressed by powers of the Planck scale, it is perhaps worthless to
look for direct phenomenological consequences. But they can play an
important role in cosmology as very weakly interactive light particles do.
In general, particles with these properties and masses in the
TeV range are problematic
for cosmology 
\cite{PWW,PP,CFKRR}. If they are stable, they can overclose
the Universe, and if they decay late, they can
dilute or alter the light elements abundance. For the fermionic fields
 this
problem could be solved if there is a period of inflation, which
would dilute their initial number density making them harmless.
For scalar fields, the problem is more serious since
after inflation they are usually sitting far from
their zero-temperature minimum
and they store a large amount of energy
in the associated oscillations.

These problems have been noticed before in the context of particular
supergravity \cite{CFKRR} and superstring \cite{ENQ} models, but
according to the arguments above they are truly generic in string
theory.
We examine alternative
mechanisms to solve these problems, finding that some of them could
be naturally implemented in phenomenologically interesting string
models. We find that the entropy problem is significantly alleviated
if the inflaton decays at late times giving a small reheating
temperature, and that
the cosmological bounds then favour low-energy
scenarios of baryogenesis (e.g. at the electroweak scale or below).
Finally, we point out that the dilaton and moduli superfields
may provide a dark matter candidate and, in $R$-parity violating
models, they could actually trigger low-temperature
baryogenesis \cite{CR}.

\vspace{0.6cm}

\section{Masses of the Dilaton and Moduli Superfields}
\vspace{0.2cm}

Following the standard notation \cite{peter}, the scalar potential for a
$d=4,\  N=1$ supergravity theory can be conveniently written
in $M_P$ units as
\begin{eqnarray}
V=\left[F_k F^l{\cal G}^k_l -3 e^{{\cal G}}\right]
+ \frac{1}{2}f_{\alpha\beta}^{-1}D^\alpha D^\beta\;\;,
\label{V}
\end{eqnarray}
where
\begin{eqnarray}
D^\alpha=-{\cal G}^i T^{\alpha j}_i z_j\;\;,\;\;\; F^l = e^{{\cal G}/2}
({\cal G}^{-1})^l_j {\cal G}^j
\label{FD}
\end{eqnarray}
Here $z_j$ denote the chiral fields, $T^\alpha$ are the gauge group
generators,
${\cal G}=K+\log|W|^2$ where $K$ is the K\"{a}hler potential
and $W$ the superpotential, $f_{\alpha\beta}$ are the gauge kinetic
functions
(in superstrings $f_{\alpha\beta}=S\delta_{\alpha\beta}$ up to threshold
corrections), and ${\cal G}^i$ (${\cal G}_j$) denotes
$\partial {\cal G}/\partial z_i$ ($\partial {\cal G}/\partial z^*_j$).
$F_l$ and $D^\alpha$ are the $F$ and $D$ auxiliary
fields. Supersymmetry is broken
if at least one of the $F_l$ or $D^\alpha$ fields takes a $vev$. The
scale of supersymmetry breakdown, $M_S$, is usually defined as
$M_S^2=\langle F\rangle$ or $\langle D\rangle$ (depending on the
type of breaking).\vspace{0.4cm}

Let us now see why the masses of the dilaton and moduli
fields are of order $m_{3/2}$. The assumptions to show this are the
following
\begin{description}

\item[i)] The $S$ and $T_i$ fields must take $vev's$ of the order of the
Planck scale $M_P$. This is certainly mandatory for the $S$ field,
whose $vev$ gives (up to threshold corrections) the gauge coupling
constant
at the unifying string scale ($\langle {Re}\ S \rangle=
g_{string}^{-2}$). If the model can be
interpreted as coming from higher dimensional compactification,
the size and shape of these compact dimensions are parametrized
by  $\langle T_i \rangle$ and are also expected
to be  of the order of the Planck scale.

\item[ii)] Supersymmetry is softly broken at a low scale
with vanishing cosmological constant. More precisely, the
mechanism for supersymmetry breakdown should yield a gravitino mass
$m_{3/2}\stackrel{<}{{}_\sim} 1\ {\rm TeV}\ll M_P$ (without unnatural
fine-tunings), as is required from phenomenological reasons.

\end{description}
We emphasize here that we do not assume any particular scenario
for achieving (i) and (ii). We just assume that if the observable
world is the low-energy limit of a string theory, the previous
conditions must be fulfilled. Besides points (i) and (ii), the
following well-known property of string perturbation theory
 should be added:

\begin{description}

\item[iii)] The fields $S$ and $T_i$ interact with gravitational strength
and  have a flat potential at string
tree-level which remains flat to all orders of perturbation theory
\cite{witten}. This means that the only source of masses will come from
non-perturbative effects
which on the other hand should be responsible also for the
supersymmetry breaking process. Hence, we will assume these fields to be
massless in the absence of  supersymmetry
breaking\footnote{It is conceivable that non-perturbative effects
which trigger masses for the $S$ and $T_i$ fields
 without breaking
supersymmetry  could exist.
However,
supersymmetry has to be
broken in any realistic model and non-perturbative effects would
 be responsible for that. The  way to relax this assumption is the
existence of a hierachy of non-perturbative effects:
Planck scale effects which fix {\it all} the $vev$'s without breaking
 supersymmetry
and low-energy effects which  break supersymmetry, as suggested
for the dilaton in \cite{FILQ}.}.

\end{description}
Once the assumptions are written, it is straightforward to find the order
of magnitude
of the masses. Roughly speaking,  (see for instance {\cite{joe}})
 after supersymmetry breaking, the mass splittings in a given
supermultiplet are given by the product of the square of the supersymmetry
breaking scale times the coupling strength between that field and
the goldstino ($\sim M_S^2/M_P$) which is just the gravitino mass.
Let us see
this more explicitly\footnote{Notice that   $S$ and $T_i$
share the properties of  the hidden sector fields in the general
analysis  of reference \cite{SW}  in the context of $N=1$ supergravity,
although
 that analysis includes
also cases such as linear superpotentials of the Polonyi  type which do not
hold
 for the moduli and dilaton fields.
It is straightforward to
prove our result using that formalism also. }.

For simplicity of notation, let us collectively denote $\phi$ the
scalar components of the $S$ and $T_i$ fields and $z$ the observable fields.
 Since the
$vev$'s of the $\phi$ fields are of order the Planck scale it is
convenient to define the dimensionless
fields $\chi \equiv \phi/M_P$. The K\"{a}hler potential can be written
as $K=M_P^2 K_0(\chi) + K_1(\chi)z^*z+\cdots$ where $K_0$ and $K_1$
are arbitrary real functions.
The important point here is that $K$ is of order $M_P^2$
and not higher, since the $\phi$ fields
have only gravitational strength interactions. Hence
${\cal G}^k_l=K^k_l\stackrel{<}{{}_\sim} O(1)$  in
(\ref{V}).

In flat space the gravitino mass is given by \cite{peter}
\begin{eqnarray}
m_{3/2}^2=e^{{\cal G}}=e^K|W|^2
\label{m32}
\end{eqnarray}
{}From (\ref{V}), (\ref{m32}) and the cancellation
of the cosmological  constant, it is clear that
\begin{eqnarray}
m_{3/2}^2\sim M_S^4 M_P^{-2}
\label{m322}
\end{eqnarray}
 Hence $V$, as given in eq.(\ref{V}),
is a sum of terms
\begin{eqnarray}
V=\sum_a{\cal V}_a
\label{VVa}
\end{eqnarray}
with
%
$\langle{\cal V}_a\rangle\ \stackrel{<}{{}_\sim}\ m_{3/2}^2M_P^2\;\;
(=M_S^4)$. Some ${\cal V}_a$ can be $\langle{\cal V}_a\rangle=0$ (or
$\ll m_{3/2}^2M_P^2$), but never $\langle {\cal V}_a\rangle\gg
m_{3/2}^2M_P^2$,
since this would imply a fine-tuning in order to cancel the cosmological
constant. Of course, $\langle{\cal V}_a\rangle=0$ in the absence of
supersymmetry breaking.

 The $\phi$ masses are then
 given by
\begin{eqnarray}
m_\phi^2=\langle\partial^2V/\partial
 \phi^2\rangle=\sum_a\langle\frac{\partial^2{\cal V}_a}
{\partial \phi^2}\rangle=\frac{1}{M_P^2}\sum_a\langle\frac{\partial^2{\cal
V}_a}
{\partial \chi^2}\rangle\sim M_P^{-2} (m_{3/2}^2 M_P^2)\sim
O(m_{3/2}^2)
\label{Mass}
\end{eqnarray}
The  additional factor coming from the
normalization of the kinetic term depends on ${\cal G}^k_l$ and  is $O(1)$.
Notice that if the breaking of supersymmetry were explicit rather than
 spontaneous the same conclusion would hold as long as the terms
in the potential
are of order $M_S^4$.
\vspace{0.4cm}
\vfill\eject

\leftline{\it Dilatino and Modulino Masses}
\vspace{0.2cm}

As usual, the fermion component of the field whose $F$ (or $D$) auxiliary
field takes a $vev$ is the goldstino, which in principle is massless but
is eaten by the gravitino through the super-Higgs effect.
This field could
perfectly be the dilatino $\tilde{S}$ or one of the modulinos
$\tilde{T_i}$,
or perhaps a certain combination of them.
 For the remaining fermionic components, $\psi_i$, of the
chiral superfields $z_i$ the mass term is $[M_\psi]^{ij}\bar\psi_{Li}
\psi_{Lj}$,
where the fermionic mass matrix $M_\psi$ can be written as
$[M_\psi]^{ij} =  \sum_{n=1}^4 [M_\psi^{(n)}]^{ij}$, with
\begin{eqnarray}
& [ M_\psi^{(1)}]^{ij} & =  -e^{K/2}|W|\left\{ K^{ij} +
\frac{1}{3}K^iK^j\right\}
\nonumber \\
&[M_\psi^{(2)}]^{ij} & =  -e^{K/2}|W|\left\{\frac{ K^{i}W^j + K^jW^i}
{3~ W}-\frac{2W_iW_j}{3W^2}\right\}
\nonumber \\
&[M_\psi^{(3)}]^{ij} & =  -e^{K/2}\sqrt{\frac{W^*}{W}}\ W^{ij}
\nonumber \\
&[M_\psi^{(4)}]^{ij} & =  e^{{\cal G}/2}{\cal G}^l({\cal G}^{-1})^k_l
{\cal G}^{ij}_k
\label{Mfer}
\end{eqnarray}
where $W^i=\partial W/\partial z_i$, $K^i=\partial K/\partial z_i$.
Notice that the canonically normalized fermion fields $\hat \psi_{Ln}$
are given by $\psi_{Lj}=U^*_{jn}\Lambda_{nn}\hat \psi_{Ln}$, where
$U$ is the unitary matrix diagonalizing the $K^i_j$ matrix and
$\Lambda=diag(\lambda_n^{-1/2})$ with $\lambda_n$ the corresponding
eigenvalues. Then the mass matrix for the $\hat \psi_{Ln}$ fields is
given by
\begin{eqnarray}
\hat M = \Lambda U^+ M (\Lambda U^+)^T
\label{Mfer2}\;\;.
\end{eqnarray}
Since in general the
non-vanishing $K_j^i$ derivatives are $\stackrel{<}{{}_\sim} O(1)$,
$M$ and $\hat M$ are of the same order of magnitude. Of course, we
are interested in the case where the $\psi_i, \psi_j$ fields
are the dilatino $\tilde S$ and the modulinos $\tilde T_i$. Let us
then analyse the magnitude of each $[M_\psi^{(n)}]^{ij}$ term
individually.

The first term $[M_\psi^{(1)}]^{ij}$ is clearly of order $m_{3/2}$
($=e^{K/2}|W|$) since the possible non-vanishing $K^{ij}, K^i, K^j$
derivatives are in general $\stackrel{<}{{}_\sim} O(1)$ in Planck units.
(Notice that for normal observable fields with zero $vev$'s
these quantities
are usually vanishing.) The second term $[M_\psi^{(2)}]_{ij}$ is
$O(m_{3/2})$ for analogous reasons together with the fact that
$W^i/W\stackrel{<}{{}_\sim} O(1)$ in Planck units. To see this, note
that an F-term, $F^k$, can be written (see eq.(\ref{FD})) as
\begin{eqnarray}
F^k=e^{K/2}|W|(K^{-1})^k_j\left( K^{j} + \frac{W^j}{W}\right)
\label{Fk}
\end{eqnarray}
where the derivatives of $K$ are $\stackrel{<}{{}_\sim} O(1)$ and
$\langle F_k \rangle\stackrel{<}{{}_\sim} M_Pm_{3/2}$. Thus
 $W^i/W\stackrel{<}{{}_\sim} O(1)$. The third term
$[M_\psi^{(3)}]^{ij}$ contains the contribution to the fermion masses
which in principle is not triggered by supersymmetry breaking
(notice that this
term is not proportional to $m_{3/2}$). On the other hand, in the
absence
of supersymmetry  breaking we know  that $m_{\tilde S}=m_{\tilde T_i}=0$.
 Consequently, for the dilaton
and
the moduli, $W^{ij}$ can only be different from zero as an effect
of supersymmetry breaking. Following a  reasoning similar to that of
  estimating the value of $\partial^2 V/\partial \phi^2$, it is
easy to see that $W^{ij}=O(W)M_P^{-2}=O(m_{3/2})$, and thus
$[M_\psi^{(3)}]^{ij}=O(m_{3/2})$. Finally, $[M_\psi^{(4)}]^{ij}$
can be written as
$[M_\psi^{(4)}]^{ij}=F^kK^{ij}_k$, where
$K^{ij}_k\stackrel{<}{{}_\sim} O(1)$ in Planck units and
$\langle F^k\rangle\stackrel{<}{{}_\sim} M_Pm_{3/2}$. Therefore,
$[M_\psi^{(4)}]^{ij}=O(m_{3/2})$.

To summarize, under the assumptions (i)--(iii)
the masses of the scalar and fermionic components of the dilaton and the moduli
are $O(m_{3/2})$
\footnote{Notice that this justifies the procedure of ref.
\cite{KL} where intermediate mass hidden fields are integrated out whereas
the moduli fields are kept in the low-energy theory.}.
\vspace{0.5cm}

\leftline{\it A Class of Models}
\vspace{0.2cm}

  We  illustrate the previous results now in a class of orbifold models in
which
 all of the moduli are frozen  except for the overall
one ($T$).
The K\"{a}hler potential in Planck scale units is \cite{DKL}
\begin{eqnarray}
K(S,T,\Phi_\alpha)=K_0(S,T)+\sum_{\a} K_1^{\a}(T){|\P_\a|}^2+
O({|\P_\a|}^4)
\end{eqnarray}
where
\begin{eqnarray}
& K_0(S,T) & = -log~ Y- 3~ log(~T_R~)\nonumber\\
& K_1^\a & = (~ T_R~)^{n_\a}
\label{Orbkahl}
\end{eqnarray}
Here $Y  = S_R+\delta~ log(~ T_R~)$, where $\phi_R=\phi + \bar\phi$,
 $\phi=S,T$. Also $\delta\equiv\frac{\delta^{GS}}{4\pi^2}$ is a model
dependent constant coming from the
one-loop anomaly cancelling Green--Schwarz counterterms and $n_\alpha$
represent the modular weights of the given matter superfield and
depend on the sector (twisted or untwisted) of the corresponding
field.
For the superpotential we will take an arbitrary function
$W(S,T,\P_\a)=W^{pert}(T,\P_\a)+W^{np}(S,T,\P_a)$ where, as we know,
the perturbative part does not depend on $S$ and vanishes with
$\P_\a$, and the non-perturbative part is unknown. To write
the explicit expression for the fermion masses, we find it
convenient to work with a general $W$ which for vanishing matter
fields will be
only the non-perturbative part $W=W^{np}(S,T)$.

The mass matrix (\ref{Mfer}) for the fermionic components of the
$S$ and $T$ fields is then given by
\begin{eqnarray}
&M^{SS} & =m_{3/2}
\left[\frac{1}{Y^2}-\frac{2}{Y}\frac{W^S}{W}-(C^{SS}+\frac{1}{3}~
 B^2)\right]\nonumber\\
&M^{TS} &= -m_{3/2}
\left[
C^{ST}+\frac{B~D}{3}+
\frac{\d~B}{Y(T_R)}(1+\frac{
 3}{A})+\frac{\d~W^T}{Y^2~A~W} \right]\\
&M^{TT} & = m_{3/2}\left[E
{}~(1-\frac{2(T_R)~W^T}{A~W})-\left(
C^{TT}+\frac{D^2}{3}\right)+\frac{\d~B}{(T_R)^2}
(1-\frac{6~\d}{A~Y})\right]\nonumber \;,
\label{Orbmfer}
\end{eqnarray}
where $A\equiv 3+\d /Y$, $B\equiv W^S/W-1/Y$ , $C^{IJ}\equiv W^{IJ}/W-
W^IW^J/W^2$, $D\equiv W^T/W-A/(T_R)$ and $E=\frac{A}{(T_R)^2}
(1+\frac{\d^2}{A~Y^2})$. For the correctly
normalized fields $\tilde \phi_1,\tilde \phi_2$ the mass matrix is
given by (\ref{Mfer2}), where in our case
\begin{eqnarray}
U^+=\left(
\begin{array}{cc}
\frac{K_S^T}{\sqrt{(\lambda_1-K_S^S)^2+(K^T_S)^2}} &
\frac{\lambda_1-K_S^S}{\sqrt{(\lambda_1-K_S^S)^2+(K^T_S)^2}} \\
\frac{\lambda_2-K_T^T}{\sqrt{(\lambda_2-K_T^T)^2+(K^T_S)^2}} &
\frac{K_S^T}{\sqrt{(\lambda_2-K_T^T)^2+(K^T_S)^2}}
\end{array}
\right)
\label{Umas}
\end{eqnarray}
and $\lambda_{1,2}=\frac{1}{2}\left(\frac{1}{Y^2}+E\right)
\pm \sqrt{\frac{1}{Y^4}+E-\frac{2}{Y^2~T_R}(A-\frac{\delta^2}{Y^2})}$.
It is
 clear that the masses are of
$O(m_{3/2})$ and the precise value will depend on the non-perturbative
superpotential. Notice that if $\delta=0$ and $W^{np}\neq
W^{np}(T)$, then $M^{TT}=0$\footnote{An apparent discrepancy with
 the calculation of \cite{DIN}  in this limit is due
 to the fact that we are using the mass matrix after the super-Higgs
effect,
which is reflected in the $1/3$ factors in eqs.(\ref{Mfer}).}.
For the bosonic fields we have also calculated the masses and found
them of the same order. For simplicity we will present here only the final
 result in the limit
$\d =0$ and for the particular case that the superpotential factorizes
$W(S,T)=\Omega (S)\Gamma(T)$. In this case the physically normalized
mass matrix does not
have off-diagonal  $(S,T)$ components and its eigenvalues are, for the
dilaton
field:
\begin{eqnarray}
M_{S_{\pm}}^2 = m_{3/2}^2~S_R^2~ \left( ~ |\frac{2\O^S}{\O}+S_R~C^{SS}|^2+
|S_R~C^{SS}|^2+(2~C^{SS}~S_R~B+h.c.)\right.\nonumber\\
 \left.\pm ~ 2|S_R~\bar B~(2~C^{SS}+
S_R~C^{SS}_S )
+\frac{2~\bar\O_{\bar S}}{\O}~(\frac{\O^S}{\O}+S_R~C^{SS})|~\right),
\label{Orbmboss}
\end{eqnarray}
whereas  for the modulus we find:
\begin{eqnarray}
M_{T_{\pm}}^2    = m_{3/2}^2~\frac{T_R^2}{9}~
 \left(~|\frac{2~\G^T}{\G}+T_R~C^{TT}|^2+ |T_R~C^{TT}|^2-(2~C^{TT}~
\bar D+ h.c.)\right.
\nonumber\\
\left.\pm ~ 2~|2\frac{\bar\G_T}{\bar\G}~(\frac{\G^T}{\G}+T_R~C^{TT})-
\bar
 D~(2C^{TT}+C^{TT}_T)|~ \right).
\label{Orbmbost}
\end{eqnarray}
Notice that unlike the observable fields for which
usually one member of the supermultiplet
acquires a mass of $O(m_{3/2})$ and the other vanishes
(since their masses are protected by gauge invariance) for these
fields both members of the supermultiplet have similar masses of
$O(m_{3/2})$,
unless there is a cancellation in any of the equations above or,
as
we said before, the non-perturbative effects do not lift the
flatness of the  potentials and some of the fields remain massless.
This
is the case if the superpotential is independent of any of the fields,
as can be easily verified.

\section{Cosmological Implications}
\vspace{0.2cm}

It is natural to ask what  the physical implications of the
lightness of the dilaton and moduli sector are. First of all it is
necessary
to note that the couplings of these particles to the observable sector
are suppressed by powers of the Planck scale, so it is perhaps useless
to
look for direct phenomenological consequences
unless they happen to be very light and could mediate long range forces
\cite{mirjam}.
 However, the
dilaton and the moduli can play an important role in cosmology as very
weakly interacting light particles do.

Let us first consider the {\em fermionic} components ($\tilde S$ and
$\tilde T_i$) of these fields.
Their cosmology is very similar to that of the gravitino \cite{PP,quien}.
In particular,
if they were stable  and in the absence of inflation, they would
overclose the Universe unless their masses are in the range
$m_{\tilde S}, m_{\tilde T_i}\stackrel{<}{{}_\sim}1$ keV
\cite{PP}. A gravitino mass of this order  is
an interesting possibility which has received some attention recently
\cite{BH} but most standard phenomenological scenarios assume
$m_{3/2}\sim$ TeV, which seems to be in conflict with cosmology.
Nevertheless, these particles are actually expected to decay (unless
they were the lightest supersymmetric particles) with a rate
$\Gamma_{\tilde \phi}\sim m_{\tilde \phi}^3M_P^{-2}$ ($\phi=S,T_i$).
They should decay before
nucleosynthesis, otherwise their decay products give unacceptable
alteration of the primordial ${}^4$He and D abundances. This imposes
a lower
bound on their masses \cite{PP,quien}
\begin{eqnarray}
m_{\tilde S}, m_{\tilde T_i}>O(10)\ {\rm TeV}
\label{cotanuc}
\end{eqnarray}
This bound is not very comfortable, but may be fulfilled
since $m_{\tilde S}, m_{\tilde T_i}=O(m_{3/2})$ includes
the possibility of one or two orders of magnitude higher than
$m_{3/2}$.

The previous cosmological problems
associated with light, very weakly interacting (but decaying)
fermionic
fields are obviated if the Universe undergoes a period of inflation
that dilutes them, provided that the reheating temperature
after inflation satisfies
\begin{eqnarray}
T_{RH}\stackrel{<}{{}_\sim}10^8 \left(\frac{100~{\rm
GeV}}{m_{3/2}}\right)\ {\rm
GeV}\label{TRH}
\end{eqnarray}
in order for them not to be regenerated in large amounts.
In this case, low-temperature mechanisms for baryogenesis, which
have recently received
much attention, become strongly favoured (for a recent review see
 \cite{CKN}).
On the other hand, it is worth mentioning a mechanism \cite{CR}
in which baryogenesis is driven by the late decay of a particle such
as the gravitino or the axino.
 It is based on the possible existence of baryon-number-violating
terms in the superpotential of the form $u_L^cd_L^cd_L^c$
(for the third generation) and exploits the sources of CP violation
 appearing in supersymmetric models.
This mechanism can also be implemented for the
dilatino and modulino fields,
since they have axino-like couplings
with gravitational strength, and
could work under condition (\ref{cotanuc}) but requires
a reheating temperature $O(M_P)$ or no inflation at
all (as is the case for the gravitino).

Let us now turn to the {\em scalar} (dilaton and moduli) fields. They
present much more severe problems than their supersymmetric partners
since in general, when they start their
relevant cosmological evolution, they are sat far from their
zero-temperature
minimum and the excessive energy associated to their oscillations
around the minimum tend to be problematic.
This well known situation \cite{CFKRR} has been called in the
literature the `Polonyi problem' or the `entropy crisis' and has been
noticed in the context of particular supergravity or superstring
models \cite{ENQ}. However, according to the arguments of the previous
section, this is a truly generic problem in string theory.

The dilaton and moduli fields are expected to be initially shifted
from their zero-temperature minimum either due to the effect of
thermal fluctuations or of quantum  fluctuations \cite{GLV} during
inflation or also due to the fact that their  coupling to the inflaton
will generally modify, during inflation,  the value corresponding to
the minimum of the potential \cite{DFN}. The shift produced by these
effects may even be as large as $O(M_P)$ but its magnitude  depends on
the particular form of the scalar potential and should be estimated
case by case.

The evolution of the $\phi$ field (canonically normalized dilaton or
moduli) is determined by
\begin{eqnarray}
\ddot{\phi}+(3H+\Gamma_\phi )\dot
{\phi}+{\partial V\over \partial\phi}=0
\label{eqphi}
\end{eqnarray}
where $\Gamma_\phi$ is the $\phi$ width ($\sim m_\phi^3 M_P^{-2}$) and
$H$ is the Hubble constant. Although $V$ is not known, if we consider
just oscillations around the minimum  taking $V\simeq {1\over
2}m_\phi^2\phi^2$ one sees that the friction dominates until the time
$t_{in}\sim m_\phi^{-1}$ ($T_{in}\sim\sqrt{m_\phi M_P}$ in the case of
radiation domination), after which
the field oscillates and the density evolves as $T^3$ \cite{PWW}. Hence,
taking as $\phi_{in}$ the value of the shift in the fields
at the end of inflation (and neglecting its evolution up to $t_{in}$),
the initial density
$\rho_\phi(T_{in})\simeq m_\phi^2\phi_{in}^2$
increases with respect to the density in radiation as
\begin{eqnarray}
\frac{\rho_\phi(T)}{\rho_{rad}(T)}=
\frac{\rho_\phi(T_{in})}{\rho_{rad}(T_{in})}\frac{T_{in}}{T}
\label{roro}
\end{eqnarray}
If $\phi$ were stable $(\Gamma_\phi\sim 0$), the constraint that $\phi$
does not overclose the Universe today imposes
\begin{eqnarray}
\phi_{in}\leq
10^{-10}M_P(m_\phi/100 {\rm GeV})^{-1/4},
\label{bound}
\end{eqnarray}
which is much smaller than the typical shifts in the fields expected
due to the mechanisms mentioned above.

However, $\phi$ will generally decay. For instance, the dilaton
$S$ is coupled to all the gauge bosons of the theory through the term
${Re}\ f~ F_{\mu\nu}F^{\mu\nu} +
{Im}\ f~ F_{\mu\nu}\tilde F^{\mu\nu}$,
where $f=S+ $ threshold corrections. Also, the moduli
fields  appear in the threshold corrections to $f$ and/or in the
Yukawa couplings between the charged fields. Therefore, $\rho_\phi$
will eventually be converted in radiation.
As long as $\phi_{in}\geq 10^{-8}M_P\sqrt{m_\phi/{\rm TeV}}$ the field
$\phi$ will dominate the energy density at the moment it decays, that
will correspond to a temperature $T_D\approx m_\phi^{11/6}\phi_{in}^{-2/3}
M_P^{-1/6}$ and will reheat the Universe to $T_{RH}$ given by
\begin{eqnarray}
T_{RH}\sim m_\phi^{3/2} M_P^{-1/2} \;\;.
\label{TR}
\end{eqnarray}
(Notice that $T_{RH}$ is essentially independent of $\phi_{in}$, provided
the Universe has been $\phi$-dominated and the total density is
$\sim\rho_c$.) Now, the decay products of $\phi$ will destroy the
${}^4He$ and $D$ nuclei, and thus the successful nucleosynthesis
predictions,
unless $T_{RH}>1$ MeV, since then the nucleosynthesis process
can be re-created. This implies
\begin{eqnarray}
m_{\phi}>O(10)\ {\rm TeV}
\label{cotanuc2}
\end{eqnarray}
This bound is, of course, similar to that of eq.(\ref{cotanuc}) for
the dilatino and modulinos (and for the gravitino),
but however it cannot be escaped with the help of
inflation. On the other hand, the entropy increase when $\phi$ decays,
$\Delta$, is given by
\begin{eqnarray}
\Delta=\left(\frac{T_{RH}}{T_{D}}\right)^3\sim
{\phi_{in}^2\over m_\phi M_P}
\label{Delta}
\end{eqnarray}
If $\Delta$ is very large, it would erase any pre-existing baryon
asymmetry.
Denoting $\Delta_{max}$ the maximum tolerable entropy production
(this will depend on the specific baryogenesis mechanism, but
$\Delta_{max}\sim 10^5$ might be acceptable) we obtain the constraint
\begin{eqnarray}
\phi_{in}^2<\Delta_{max}M_P m_\phi
\label{Delta2}
\end{eqnarray}
We want to note here that a possible way out of this dilema would be
to have the baryogenesis generated by the $\phi$ decays,
that would
take place just before nucleosynthesis.
This could be
done for instance if the $\phi$ decay into gaugino pairs is allowed
(which seems reasonable in view of (\ref{cotanuc2})) and implementing
then a mechanism similar to that of ref.\cite{CR} just discussed.
Unlike the baryogenesis scenarios
based on decays of fermions (gravitino, dilatino, etc.)
those with the $\phi$ decays would not imply any severe constraint on
the
reheating after inflation since it is easier to dominate the energy
density
at the moment of decay.

This situation can however be ameliorated: the previous discussion
assumed that at the moment in which $\phi$ starts to oscillate, the
inflaton has already decayed and produced the main reheating of the
Universe. However, this is not necessary, especially in view of the
many scenarios of low-temperature baryogenesis being considered
nowadays. If the inflaton $\varphi$ were instead to
decay at $t_{RH}\gg t_{in}$,
the situation will significantly change due to the fact that between
those times both the inflaton and $\phi$ will oscillate and their
energies will both evolve as $a^{-3}$, where $a$ is the scale factor.
Hence, using the fact that, at $t_{in}$, $H=m_\phi$ so that
$\rho_\varphi=3 M_P^2m_\phi^2/(8\pi)$, we have
\begin{eqnarray}
{\rho_\phi\over\rho_\varphi}(t_{RH})\simeq {8\pi\phi_{in}^2\over
3M_P^2}\;\;.
\end{eqnarray}
Only after the inflaton energy is converted into radiation
($\rho_\varphi(t_{RH})\sim T_{RH}^4$) the relative contribution of
$\phi$ to the density starts to increase as
\begin{eqnarray}
{\rho_\phi\over\rho_\varphi}(T)\simeq {8\pi\phi_{in}^2\over
3M_P^2}{T_{RH}\over T}\;\;
\end{eqnarray}
and, for a stable $\phi$, the condition that at $T=3$ K the Universe
is not overclosed becomes
\begin{eqnarray}
\phi_{in}\leq\sqrt{1 {\rm TeV}\over T_{RH}}10^{-6}M_P
\;\;,
\label{stable}
\end{eqnarray}
that is much less severe than the previous bound if the reheating
temperature is of the order of the weak scale.

In the more plausible situation  in which $\phi$ decays, by a similar
reasoning we obtain that $\phi$ will be dominating the density of the
Universe at the decay time only if
\begin{eqnarray}
\phi_{in}\geq\left({m_\phi\over 10\ {\rm TeV}}\right)^{3/4}\left(
{T_{RH}\over 100\ {\rm GeV}}\right)^{-1/2}\ 10^{-3} M_P
\;\;,
\end{eqnarray}
and in that case the Universe has a further reheating to
$T_{RH}'\simeq \sqrt{m_\phi^3/M}$ with an increase in the entropy by a
fraction
\begin{eqnarray}
\Delta\simeq{\phi_{in}^2T_{RH}\over (M_Pm_\phi)^{3/2}}
\;\;.
\end{eqnarray}
The analogue of eq. (\ref{Delta2}) is
%
$\phi_{in}^2<\Delta_{max} (M_P m_\phi)^{3/2}/T_{RH}$
%
relaxing the bounds for small enough values of $T_{RH}$.
For the indicative values used before (remember that $m_\phi>10$
TeV eliminates the problems associated to nucleosynthesis and
$T_{RH}\sim 100$ GeV could allow for electroweak baryogenesis to occur)
$\Delta$ is still not unreasonably large even for $\phi_{in}=O(M_P)$
($\Delta \leq 10^5$), although clearly for those
values of $\phi_{in}$ the approximation
of a quadratic potential goes badly wrong and the detailed form of the
potential would have to be taken into account.

Hence, although the problem associated to the scalar field
oscillations is severe, we have seen that there are ways to alleviate
it. Furthermore, in the context of supergravity models,
there are two interesting proposals to solve this `entropy
crisis' \cite{DFN,OS}. The first method \cite{DFN} is based on a
scenario in which supersymmetry is broken
through an O'Raifeartaigh superpotential. Then, the scalar
field  has almost vanishing $vev$ and a
rather large mass ($\sim (m_{3/2}M_P)^{1/2}$) at zero-temperature,
thus the contributions to the scalar potential during inflation do not
 change
appreciably the position of the minimum. Consequently, the energy
stored in the scalar field is very small, avoiding cosmological
problems. Besides the fact that this mechanism
does not address the problem of quantum fluctuations, it
  cannot be
implemented for the $S$ and $T$ fields since they must have $O(M_P)$
(non-vanishing) zero-temperature $vev$'s and their masses should be
$O(m_{3/2})$.

 A further mechanism that has been proposed \cite{OS} is to
couple the problematic fields to heavy ones that decay
promptly into radiation. In this way the coupled evolution of the
scalar fields may allow for a transfer of the energy stored
in the oscillations to the radiation. This method might be accomodated
in the context of string
theories for the $S$ and $T$ field since there are heavy  fields
with mass terms in the perturbative superpotential that, in
general, will depend on the moduli fields. Likewise, the factor $e^K$
in front of the scalar potential provides a non-suppressed coupling
between the dilaton and the massive fields.
This simply reflects the fact that all the physical couplings are
proportional to the string coupling constant. There is still an
additional source of couplings between the dilaton and massive
fields. It is known that if the four-dimensional gauge group of the
string has an `anomalous' $U(1)$ factor, a Fayet--Iliopoulos $D$-term
is generated in string  perturbation theory \cite{DSW}. The
corresponding
term in the scalar potential has the form
\begin{eqnarray}
V_{FI}\sim\frac{1}{S_R}\left| \frac{Tr\
Q^{a}}{48\pi^2}\frac{1}{S_R}\ + \ \sum_i Q^a_i |C_i|^2
\right|^2
\label{FI}
\end{eqnarray}
where $Q^a_i$ are the anomalous charges of the scalar fields $C_i$.
Thus, there appear effective mass terms of the form $\sim (S+\bar
S)^{-2} |C_i|^2$. 
Furthermore, in this
scenario there appear masses for the dilatino and
${Im} \ S$, through their couplings to the `anomalous' photino
and photon respectively, thus the cosmological problems associated
with these fields seem to be reduced
(although  there remains a combination of dilaton and matter fields
which remains massless at this level and to which the previous analysis
applies). It is  quite
suggestive that most of the phenomenologically interesting
superstring models have been constructed within this framework
\cite{CM}. For the remaining fields, the possibility
that the transfer of energy mentioned
previously be efficient would be worth  of a detailed analysis,
although that is beyond the scope of the present paper.

Let us also mention that
another danger coming from the dilaton and moduli fields being initially
sat far from the zero-temperature minimum is that they might
never fall into it if there is a barrier in between. (This problem has
been recently stressed in ref.\cite{BS}.) This can typically happen for the
real part of the dilaton (${Re}\ S$) if it initially sits at ${Re}\
S\gg{Re}\ S_o$, since for ${Re}\ S\rightarrow \infty$ all the gauge and
gravitational interactions are switched off,
and thus supersymmetry is restored and $V\rightarrow 0$. Therefore it is
necessary to assume that the initial value of the generic scalar field,
$\phi_{in}$, is basically in the slope leading to the zero-temperature
minimum.
A further problem is that for the typical potentials arising from
gaugino condensation, the kinetic energy of the dilaton may be so large
as to prevent inflation to take place \cite{BS}, unless some mechanism
exist to push $\phi$ close to the minimum (this problem may be
only
characteristic of those potentials and not necessarily generic in
string theory).

Finally we will mention that the dilaton and moduli-sector fields may
also provide candidates for dark matter. As is apparent from the
previous discussion, if the fermionic fields $\tilde \phi$ were
stable they would close the Universe for $m_{\tilde \phi}\sim 1$ keV
in the absence of inflation, or if the $\tilde \phi$ mass is set to a
$T_{RH}$-dependent value if there is an
inflationary epoch, in a similar way as
for the gravitino \cite{MMY}
(for instance $m_\phi\simeq 100$ GeV for $T_{RH}\simeq 10^9$ GeV).
If some of the scalar fields $\phi$ were stable,
they would close the Universe if the bounds in eqs. (\ref{bound})
or (\ref{stable}) are
saturated. In that case,
the missing energy of the Universe would be stored in the coherent
oscillations of the $\phi$-field in a similar way to the more
standard axionic dark matter.

\vspace{0.6cm}

We would like to acknowledge useful conversations with C. Burgess,
L. Ib\'a\~nez, J. Louis, S. Mollerach and M. Quir\'os.
After this work was finished, we received a preprint of T. Banks,
D. Kaplan and A. Nelson where the cosmological problems associated
with dynamical supersymmetry breaking are addressed.

\small


\begin{thebibliography}{99}




\newcommand{\bi}[1]{\bibitem{#1}}

\bi{HMV} J. Harvey, G. Moore and C. Vafa, Nucl. Phys. B304 (1988) 269.

\bi{AMQ}I. Antoniadis, Phys. Lett. B246 (1990) 377\\
I. Antoniadis, C. Mu\~noz and M. Quir\'os, Nucl. Phys. B297
(1993) 515.

\bi{PWW} J. Preskill, M.B. Wise and F. Wilczek, Phys. Lett. B120 (1983)
127;\\
L.P. Abbot and P. Sikivie, Phys. Lett. B120 (1983) 133;\\
M. Dine and W. Fischler, Phys. Lett. B120 (1983) 137.


\bi{PP} H. Pagels and J.R. Primack, Phys. Rev. Lett. 48 (1982) 223;\\
S. Weinberg, Phys. Rev. Lett. 48 (1982) 1303.


\bi{CFKRR} G.D. Coughlan, W. Fischler, E.W. Kolb, S. Raby and G.G. Ross,
 Phys.
 Lett. B131 (1983) 59.

\bi{ENQ} J. Ellis, D. Nanopoulos and M. Quir\'os, Phys. Lett. B174 (1986)
 176.

\bi{CR} A. Cline and S. Raby, Phys. Rev. D43 (1991) 1781\\
S. Mollerach and E. Roulet, Phys. Lett. B281 (1992) 303.

\bi{peter} E. Cremmer, S. Ferrara, L. Girardello and A. Van Proyen,
Nucl. Phys. B212 (1983) 413;
H.P. Nilles, Phys. Rep. 110 (1984) 1.

\bi{witten} E. Witten, Phys. Lett. B155 (1985) 151; Nucl. Phys. B268
(1986) 79; \\
C. Burgess, A. Font and F. Quevedo, Nucl. Phys. B272 (1986) 661;\\
L. Dixon in {\it Superstrings, Unified Theories and Cosmology},
B. Furlan $et$ $al.$ eds. (World Scientific, 1988);\\
M. Dine and N. Seiberg, Phys. Rev. Lett. 57 (1986) 2625.

\bi{FILQ} A. Font, D. L\"ust, L.E. Ib\'a\~nez and F. Quevedo, Phys. lett. B249
 (1990) 35.

\bi{joe} J. Polchinski, {\it Introduction to Supersymmetry} SLAC
summer Institute (1985) 1.

\bi{SW} S.K. Soni and A. Weldon, Phys. Lett. B126 (1983) 215.


\bi{DKL} L. Dixon, V. Kaplunovsky and J. Louis, Nucl. Phys. B329 (1990) 27;\\
J.--P. Derendinger, S. Ferrara, C. Kounnas and F. Zwirner, Nucl. Phys.
B372 (1992) 145.

\bi{KL} V. Kaplunovsky and J. Louis, Phys. Lett. B306 (1993) 269.


\bi{DIN} J.--P. Derendinger, L.E. Ib\'a\~nez and H.P. Nilles,
 Nucl. Phys. B267
 (1986) 365.
\bi{mirjam} M. Cveti\v c, Phys. Lett. B229 (1989) 41.

\bi{quien} M. Yu. Khlopov and A.D. Linde Phys. Lett. B138 (1984) 265;\\
J. Ellis, J.E. Kim and D. Nanopoulos, Phys. Lett. B145 (1984) 181;\\
J. Ellis, D. Nanopoulos and S. Sarkar, Nucl. Phys. B259 (1985) 175.




\bi{BH} C.P. Burgess and O. Hernandez, McGill preprint (1993)\\
M. Dine and A. Nelson, UCSD preprint (1993)\\
E.J. Chung, H.B. Kim and J.E. Kim, Seoul preprint SNUTP--93--22 (1993).

\bi{CKN} A. Cohen, D. Kaplan and A. Nelson, UCSD preprint (1993).



\bi{GLV} A.S. Goncharev, A.D. Linde and M.I. Vysotsky, Phys. Lett.
B147
(1984) 279.



\bi{DFN} M. Dine, W. Fischler and D. Nemechansky, Phys. Lett. B136
(1984)
169\\
G.D. Coughlan, R. Holman, P. Ramond and G.G. Ross, Phys. Lett. B140
(1984) 44.

\bi{OS} B. Ovrut and P.J. Steinhardt, Phys. Lett. B147 (1984) 263;\\
O. Bertolami, Phys. Lett. B209 (1988) 277.



\bi{DSW} M. Dine, N. Seiberg and E. Witten, Nucl. Phys. B289 (1987)
589.


\bi{CM}L.E. Ib\'a\~nez, J.E. Kim, H.P. Nilles and F. Quevedo, Phys.
Lett. B191 (1987) 282;\\
 J.A. Casas and C. Mu\~noz, Phys. Lett. B214 (1988) 63;\\
A. Font, L.E. Ib\'a\~nez, H.P. Nilles and F. Quevedo, Phys. Lett.
B210 (1988) 101;\\
I. Antoniadis, J. Ellis, J. Hagelin and D. Nanopoulos, Phys. Lett. B205
(1988) 459.

\bi{BS} R. Brustein and P.J. Steinhardt, Phys. Lett. B302 (1993) 196.

\bi{MMY} T. Moroi, H. Murayama and M. Yamaguchi, Phys. Lett. B303
(1993) 289.


\end{thebibliography}
\end{document}